\documentclass[10pt]{iopart}

\usepackage{amssymb}
\usepackage{graphicx}
\usepackage{epstopdf}
\usepackage{textcomp}
\usepackage{color}
\usepackage{filecontents}
\usepackage{cite}
\usepackage{url}
\usepackage{iopams} 
\expandafter\let\csname equation*\endcsname\relax
\expandafter\let\csname endequation*\endcsname\relax 
\usepackage{amsmath}
\usepackage{amsfonts} 

\begin{document}

\title[Confined H]{Confined H(1s) and H(2p) under different geometries.}
\author{G. Micca Longo$^1$ , S. Longo$^{1,2,3}$ and D. Giordano$^4$}

\address{$^1$ CNR-Nanotec, via Amendola 122/D, 70126 Bari, Italy}
\address{$^2$ Department of Chemistry, University of Bari, via Orabona 4, 70126 Bari, Italy}
\address{$^3$ INAF-Osservatorio Astrofisico di Arcetri, Largo E. Fermi 5,
              I-50125 Firenze, Italy}
\address{$^4$ ESA-ESTEC, Aerothermodynamics Section, Kepleerlaan 1, 2200 ag, Noordwijk,
              The Netherlands}

\ead{gaia.miccalongo@nanotec.cnr.it}
\ead{savino.longo@nanotec.cnr.it}
\ead{domenico.giordano@esa.int}

\vspace{10pt}
\begin{indented}
\item[]February 2014
\end{indented}

\begin{abstract}

In this paper the Diffusion Monte Carlo (DMC) method is applied to the confined hydrogen atom with different confinement geometries. This approach is validated using the much studied spherical and cylindrical confinements and then applied to cubical and squared ones, for which data are not available, as new applications of the method relevant to solid state physics. The energy eigenvalues of the ground state and one low-lying excited state are reported as a function of the characteristic confinement length. 

\end{abstract}

\noindent{\it Keywords}: Confined H, DMC, Lyman-$\alpha$


\section{Introduction}
\label{intro}
The problem of the confined hydrogen atom has been the subject of many studies in the past \cite{michels1937,degroot1946,yngve1988,yngve1986,aquino2007},
in view of its importance as a basic theoretical problem and as a model for systems in 
astrophysics (interior of stars), geophysics (giant planets), nuclear science (compression techniques for nuclear fusion).

Early attempts trace 
back in time to the works of Michels \emph{et al.}\cite{michels1937} and De Groot and Ten Seldam \cite{degroot1946} in their attempt to 
investigate the dependence of hydrogen polarizability on extreme pressure conditions. 
These researchers considered a modification of the boundary conditions of the 
quantum problem: the usual condition associated with the unconfined atom, 
i.e. the hydrogen-atom internal wave function must vanish at infinity, was replaced with the condition 
that the wave function must vanish, or have a node, at a finite distance $r=r_0$.

It was then recognized that a better understanding of the interiors 
of planets like Jupiter and Saturn depends upon the knowledge of the properties of atomic hydrogen 
at high pressures \cite{pfaffenzeller1996,dutt2001}.
More recently, Capitelli and Giordano \cite{capitelli2009} considered the problem in the context of 
aerothermodynamics. 
Besides high-pressure situations, confined hydrogen atoms can be present inside metal-alloy lattices at
normal pressure. 

Atoms imprisoned in zeolite traps, clusters and in fullerene cages provide a very good examples of confined systems in the atomic physics and inorganic chemistry \cite{soullard2004,connerade1999,belosludov2003,
decleva1999,hernandez1998,qiao2002,connerade2001}.

The model of the spherical confined hydrogen atom has been applied to the semiconductor physics. In particular, the hydrogen-like donors confined in quantum dots are well described by this model \cite{kang2007}. In fact, the quantum dot Hamiltonian is similar to the confined atom one, except that the system is enclosed in a harmonic oscillator potential.

These new applications, and some of those reviewed before, ask for confinements 
with more complex shapes than those considered in previous studies to account for reduced simmetry due
to the crystal lattice, such as four-fold ($T_d$) and six-fold ($O_h$) coordination.
An example
is given by the storage of hydrogen into solid materials to be used as energy
source for transportation. In this respect, theoretical and experimental
results show that hydrogen in atomic form is confined into octahedral
and tetrahedral cavities of metals (Mg, Ti), metal clusters, alloys and
inter-metallic compounds (e.g. $LaNi_5$) \cite{cremaschi1985,sakaguchi1995,schlapbach2001}. 
Moreover, polarizability studies under strong electric field require
consideration of several values of proton displacement from the center
of the cavity, something that has been considered only for a few cases so far \cite{ting2000}.

In view of the perspectives opened by such a wide range of studies and applications,
a methodology is required to produce results for the ground state and the lowest excited states 
of the hydrogen atom confined within
cavities of complex shape and with the possibility to displace the nucleus easily. Such a procedure
can be formulated using the diffusion Monte Carlo (DMC) method, which operates for this class of problems under the
most favorable conditions, by allowing to consider confining walls of any geometry while retaining the
use of cartesian coordinates and to include any distribution of positive charge within the cavity.
Low lying excited states can be considered by including appropriate symmetry-based nodal surfaces, an approach demonstrated for H$_2^+$ states for confined systems in a previous paper by the present authors \cite{micca2015}.

In this paper, the method is illustrated by reproducing the results for
ground and low-lying excited states obtained in previous studies, and then by reporting new results for other geometries.  All the
results are presented in length (bohr, $a_0$) and energy (hartree, $Ha$) atomic units.  \\

\section{Diffusion Monte Carlo method}\label{DMC}
The Quantum Monte Carlo (QMC) methods are powerful and general tools for computing the electronic ground state of atoms, molecules and solids. 
In particular, the Diffusion Monte Carlo (DMC) method is a stochastic method that makes use of Green's function for evolving a solution of the imaginary-time 
Schr\"{o}dinger equation. Ideally, DMC is an exact numerical method but in practice numerical parameters are used in order to control the computational 
simulations.
In our simulations, the hydrogen atom is placed inside an impenetrable spherical box and the proton is assumed fixed at the center of the box. 
The electron is replaced by a chain of fictitious particles called \emph{walkers} \cite{foulkes2001}; the number \emph{N} of walkers becomes a parameter of the 
simulation. The Coulomb potential acts on every walker.

During the simulation, the walkers diffuse throughout the phase space and the transition probability density for the evolution of the walkers is given by the approximate Green's function:

\begin{equation}\begin{split} \label{1}
G(\textbf{R}\leftarrow\textbf{R'},\tau)\approx \left(2\pi\tau\right)^{-\frac{3N}{2}}exp\left[-\frac{\left(\textbf{R}-\textbf{R'}\right)^2}{2\tau}\right]\\
exp\left[-\frac{\tau\left(V\left(\textbf{R}\right)-V\left(\textbf{R'}\right)-2E_T\right)}{2}\right]
\end{split}\end{equation}

Every diffusion step consists in two phases: propagation and branching. At the beginning, 
each walker is moved from its old position \textbf{R} to the new one \textbf{R'} by the probability

\begin{equation}\label{1}
T=\left(2\pi\tau\right)^{-\frac{3N}{2}}exp\left[-\frac{\left(\textbf{R}-\textbf{R'}\right)^2}{2\tau}\right]
\end{equation}\\
where $\tau$ is the time step.

The factor that determines the number of walkers surviving for next step is given by

\begin{equation}\label{2}
P=exp\left[-\frac{\tau\left(V\left(\textbf{R}\right)-V\left(\textbf{R'}\right)-2E_T\right)}{2}\right]
\end{equation}\\
in which $V$ is the potential energy and $E_T$ is the so called energy offset that controls the total population of the walkers. When $P<1$, the walker continues its evolution with 
probability $P$ and dies with probability $1-P$; when $P\geq1$, the walker continues its evolution and, at the same position, a new walker is created with probability $P-1$. From equation (\ref{2}) it is clear that the walkers tend to proliferate in regions of low potential energy and to disappear from regions of high potential energy.

The energy offset $E_T$ is determined by keeping track of changing walkers and by tuning it at every step in order to make the average walker population approximately constant \cite{thijssen}. A simple formula for adjusting $E_T$ is

\begin{equation}\label{3}
E_{T_i}=E_{T_{i-1}}+\tau\ln\left(\frac{N_{i-1}}{N_i}\right)
\end{equation}
where $E_{T_{i-1}}$ is the energy value at time step $i-1$ and ${N_{i-1}}$ and $N_i$ are respectively the number of walkers at time step $i-1$ and $i$.

A time average has been used in order to reduce the fluctuations and improve the energy calculations. 

In view of the low dimensionality of the present problem, even a straightforward approach that does not use importance sampling \cite{reynolds1982,foulkes2001} may produce satisfactory results. This has been checked by direct comparison with calculations that make use of an importance sampling transformation with 
trial wavefunction of the form\\

$\psi_T=exp(-r)\frac{sin(\pi r/r_0)}{r}$ .\\ 

This trial wavefunction takes into account the peculiarities of the present system: the singularity at the origin ($exp(-r)$) and the boundary conditions associated with the presence of the barrier (${sin(\pi r/r_0)}{r}$). 

The use of importance sampling did not decisively affect the computational cost of results stable enough to be used in the plots reported, which have been therefore calculated without using importance sampling.\\

Although DMC is principally a ground state method, it can still be used for low excited states with distinct symmetry. 
The starting point is the fixed-node DMC method \cite{anderson1975,anderson1976,moskowitz1982,reynolds1982}.
The basic idea of the fixed-node DMC method is very simple: a trial wave function is chosen and used to define a trial node surface that reproduces 
the symmetries of the excited state.
 
Here the $2p$ state ($n=2$, $l=1$) eigenvalues are reproduced. An extra absorbing boundary is introduced at the centre of the box, along the x direction, 
in order to represent the symmetry of this particular state.
 
Thus, a DMC code has been written consistently with the preceding description and used to produce results discussed in the following sections.

\section{Spherical and circular confinement}\label{valid}
\begin{figure}
\centering
\resizebox{0.5\textwidth}{!}{
  \includegraphics{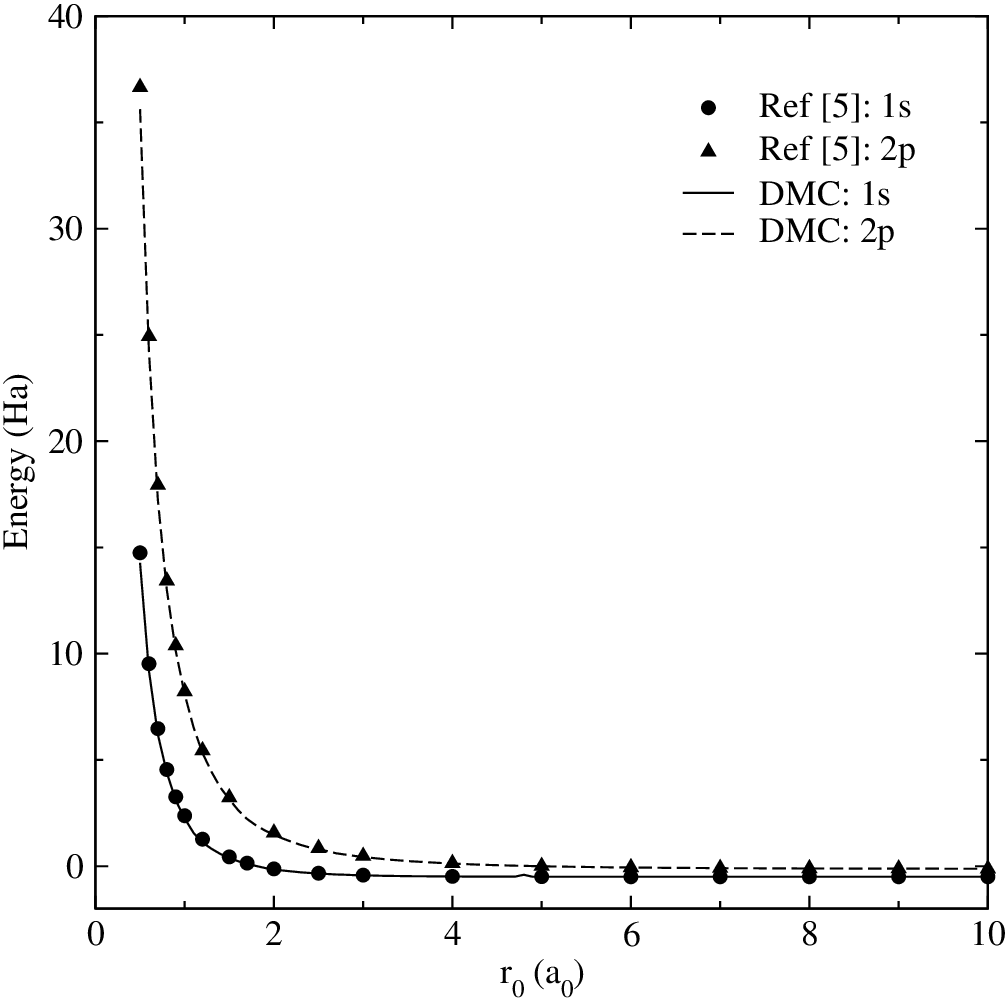}
}
\caption{Confined hydrogen atom: spherical box. Comparison between Aquino \emph{et al.} \cite{aquino2007} data sets and DMC ones.}
\label{fig:1} 
\end{figure}

Figure \ref{fig:1} shows the ground state and the $2p$ state energies of the three-dimensional confined hydrogen atom as a function of box radius $r_0$, as obtained by DMC code and those published 
by Aquino \emph{et al.} \cite{aquino2007}.

About $10^3$ walkers have been used in the simulations, while $\tau$ values ranged between $10^{-2}$ for free cases and weak confinement and $10^{-4}$ for strong confinements and excited state.

The DMC method is able to reproduce the best data with good 
accuracy, in spite of its simple implementation in cartesian coordinates. 
Small systematic difference with the best data is typically negative since 
the sphere appears larger than $r_0$ to walkers; the difference can be 
further reduced by reducing $\tau$, but the computational cost increases considerably.\\

One of the key factors of the DMC code is its extreme flexibility with respect to a change of space dimensions. 
Equations (\ref{1}) and (\ref{2}) are rearranged accordingly, for $1s$ and $2p$ states, in order to reproduce the two-dimensional systems.

\begin{figure}
\centering
\resizebox{0.5\textwidth}{!}{
  \includegraphics{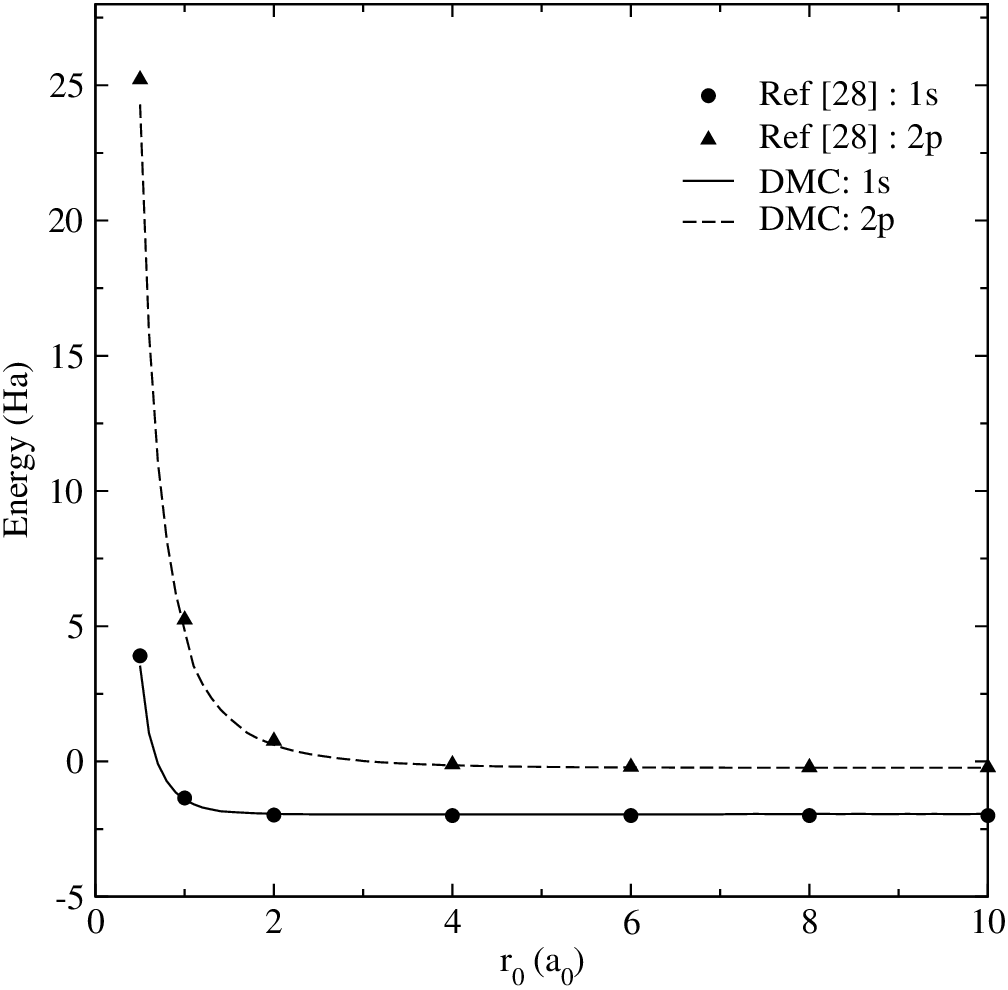}
}
\caption{Confined hydrogen atom: circle box. Comparison between Aquino \emph{et al.} \cite{aquino2005} data sets and DMC ones.}
\label{fig:2} 
\end{figure}

Figure \ref{fig:2} shows the ground state and the $2p$ state energies of the two-dimensional confined hydrogen atom as a function of box radius $r_0$, as obtained by our DMC code and compared with those published 
by Aquino \emph{et al.} \cite{aquino2005}. 

The accuracy is good also in this case and the 
considerations of the previous case apply again.

\section{Squared and cubical confinement}\label{cubic}

In the cubical confined case and in the squared one, there is no set of results to compare since previous calculations by different methods are not available. This is surprising since these confinements are potentially important for numerous applications where the corresponding symmetry of confinement is met.

The effects of a cubical and squared confinement on the hydrogen atom are studied here in order to demonstrate the DMC code efficiency with respect to a change of geometry. In this context, cubical confinement is the most simple choice ad it has the same symmetry ($O_h$) of octahedral cavities in solids. Furthermore, a cubical box is a fair approximation of the crystal repulsive field which is produced by the six ions placed in the vertexes of an octahedral cavity \cite{burns1993}.

\begin{figure}
\centering
\resizebox{0.5\textwidth}{!}{
  \includegraphics{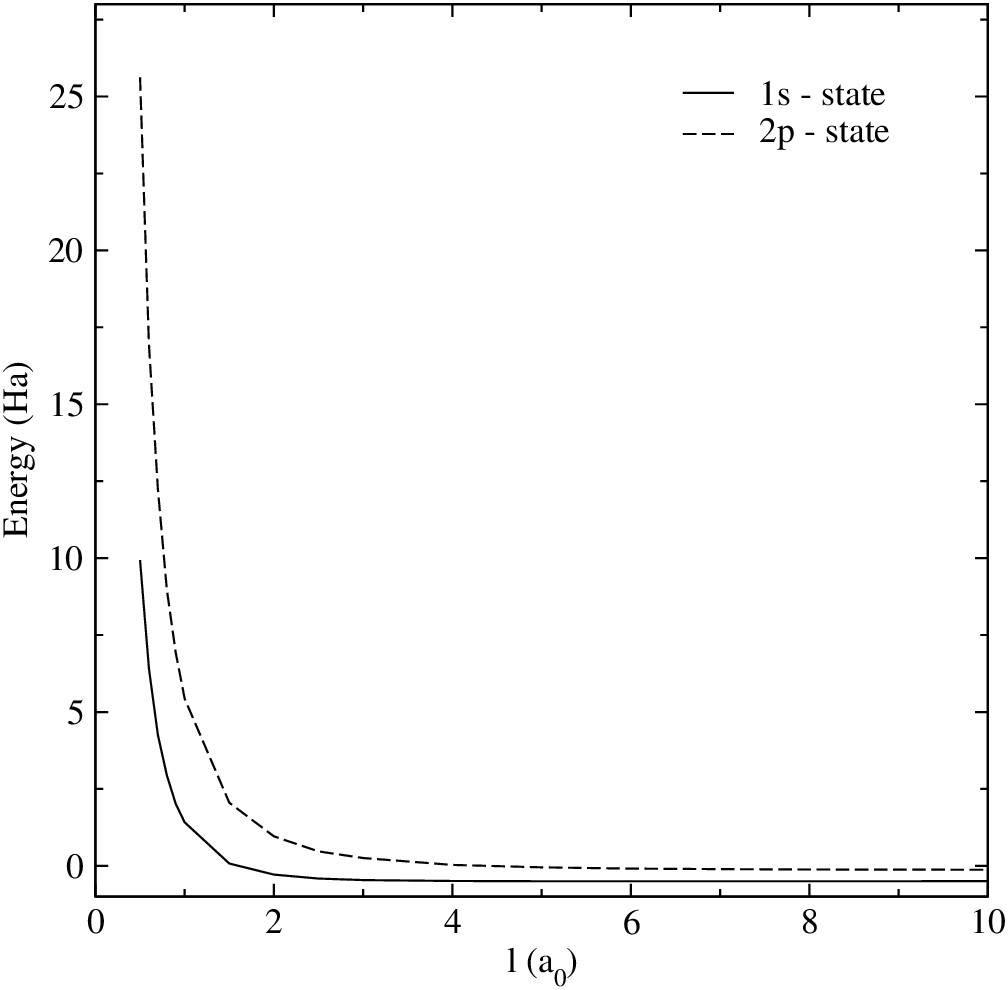}
}
\caption{Confined hydrogen atom: cubical box.}
\label{fig:3} 
\end{figure}

\begin{figure}
\centering
\resizebox{0.5\textwidth}{!}{
  \includegraphics{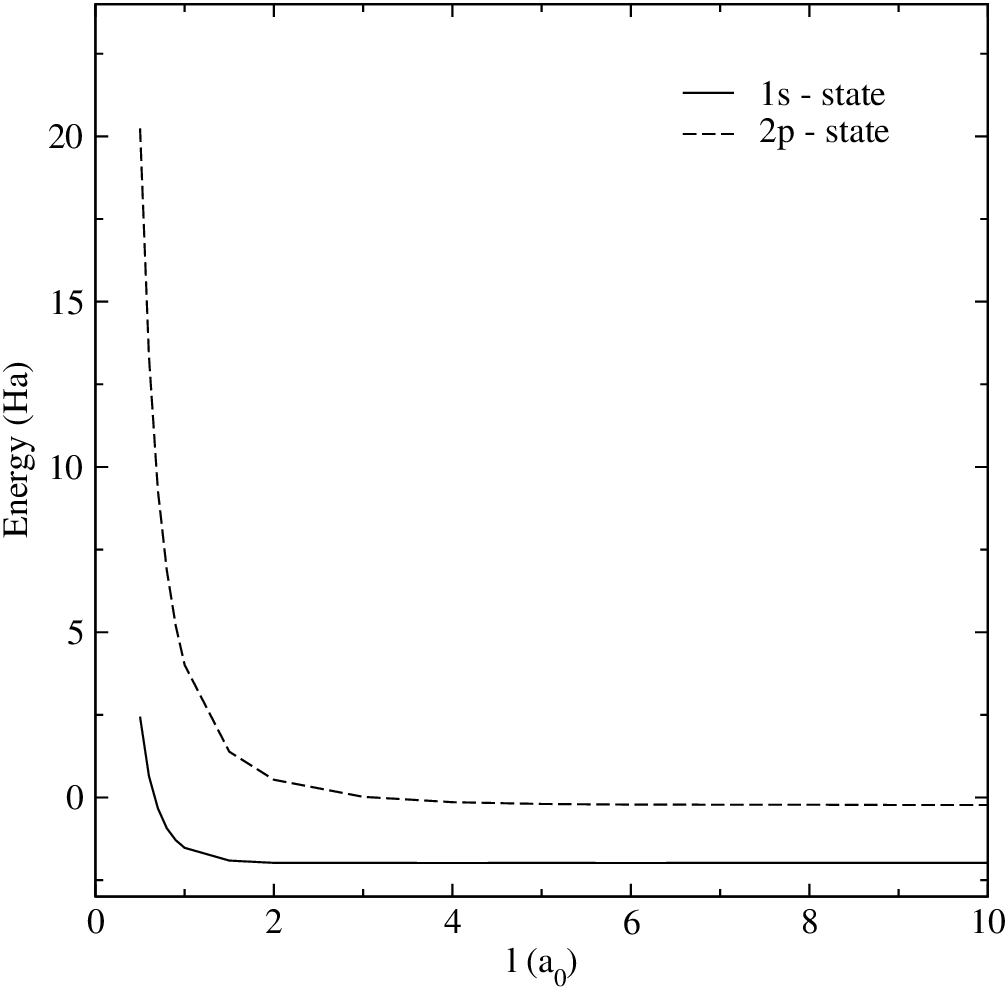}
}
\caption{Confined hydrogen atom: squared box.}
\label{fig:4} 
\end{figure}

Keeping the same code structure and cartesian coordinates, only details of the subroutine for the calculation of the potential, as a function of the walker position, are updated. 
Figures \ref{fig:3} and \ref{fig:4} show respectively the ground state and the $2p$ state energies of the hydrogen atom confined in a cubical and in a squared box, as a function of box side dimension $l$, as obtained by DMC code. For the excited state, an additional barrier confines walkers in the $x>0$ semi-space and semi-plane respectively, in order to get the appropriate symmetry.

As in the prevoius cases, about $10^3$ walkers have been used, and $\tau$ values ranged between $10^{-2}$ for free cases and weak confinement and $10^{-4}$ for strong confinements and excited state.

These results for the cubical and squared confinement are reported for the first time, and, besides providing an illustration of the proposed procedure, may be a useful reference for studies in the field reviewed above.

\section{Lyman-$\alpha$ line shift}

The confinement of H atom inside a physical cavity is expected to affect the position of line spectra, providing a perspective method to investigate these systems. Such possibility starts to be considered in actual experiments \cite{cassidy2011}. 
In this respect, the calculation of the Lyman-$\alpha$ line as a function of the confinement dimension and shape can be very interesting.

We calculate the wavelenght from the energy eigenvalues corresponding to 
the initial and final state by the usual formula:

\begin{equation}\label{lamda}
\lambda=\frac{hc}{E_i-E_f}
\end{equation}

\begin{figure}
\centering
\resizebox{0.5\textwidth}{!}{
  \includegraphics{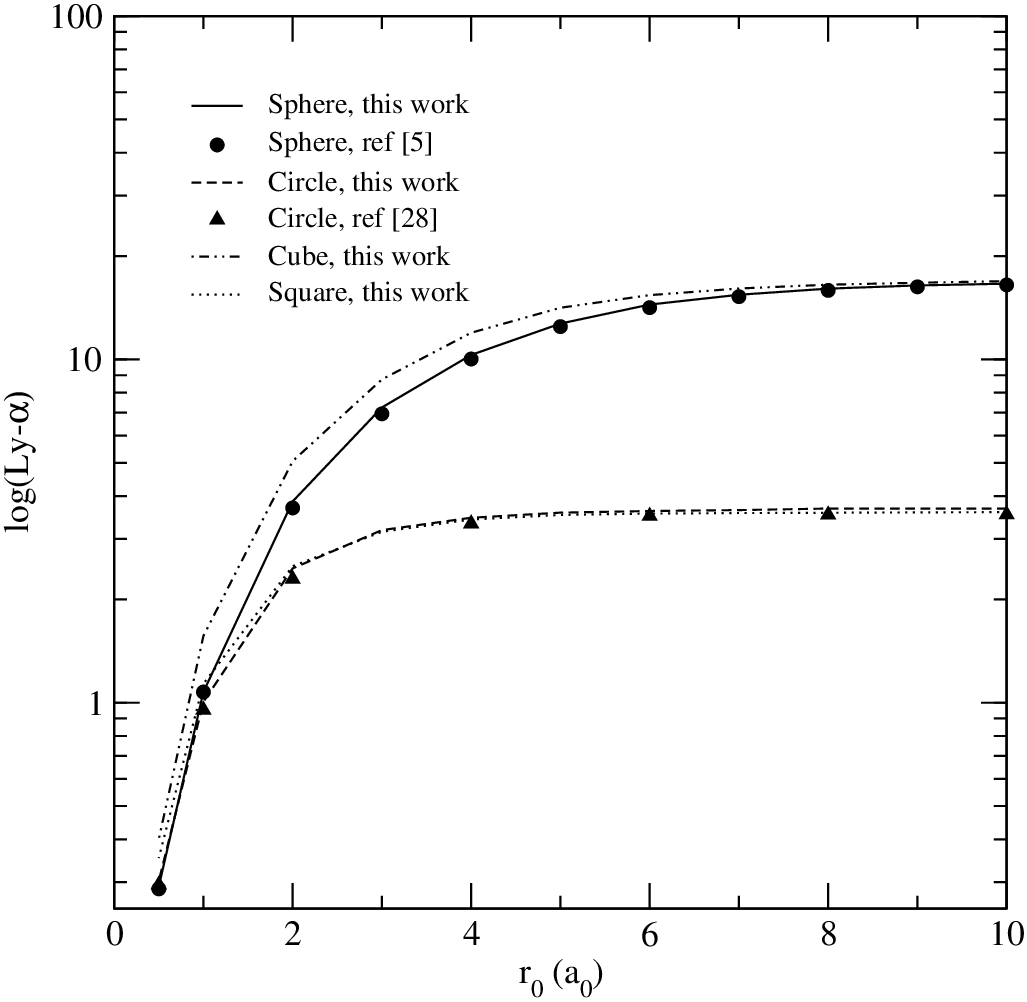}
}
\caption{Lyman-alpha line.}
\label{fig:5} 
\end{figure}

In the present paper, atomic units has been used, so $c=1$ and $h=2\pi$; the values of $E_i$ and $E_f$ are the ones used for the previous curves and given by the DMC code. Results are shown in figure \ref{fig:5}.

It can be seen that dimensionality is a factor much more important than 
detailed shape to fix the Lyman-$\alpha$ wavelenght. The difference between 
cubic and spherical confinenents is however quite important, and 
potentially accessible to experimental measurements for H atoms confined 
into solid surfaces.

\section{Conclusions}\label{conc}
The emergence of several new applications and challenging problems related to the geometric confinement of hydrogen atoms asks for a simple and flexible methodology to explore different geometries.

The DMC method with properly 
placed boundaries has the advantage of 
using cartesian coordinates in all cases and by allowing very simple and straightforward 
changes in boundary conditions and dimensionality. Excited states can be 
reproduced by using properly placed nodal surfaces.

Although DMC is not the best method to provide highly precise results like those of recent papers cited above \cite{aquino2007,aquino2005}, we believe that our method meets a demand for a simple technique to investigate variable shapes, nuclei positions and problem dimensionality. 

Complex and symmetric shapes
for the confining wall, inspired by the symmetry of cavites in solid
state matter, can extend the range of application of this inspiring,
long tradition model. Possibilities for future work based on the
proposed method include tetrahedral boxes (as a model for cavities of the same shape in solids), displaced nucleus from the center, extension to
molecular systems. \\

\section*{Acknowledgment}
This research activity has been supported by the General Studies Programme of the
European Space Agency under 
grant 4200021790CCN2.


\section*{References}

\begin{thebibliography}{}
\bibitem{michels1937} A. Michels, J. de Boer and A. Bijl, \emph{Physica} \textbf{4} (1937) 981.    
\bibitem{degroot1946} S. A. De Groot and C. D. Ten Seldam, \emph{Physica Amsterdam} \textbf{12} (1946) 669. 
\bibitem{yngve1988} S. J. Yngve, \emph{J. Math. Phys.} \textbf{29} (1988) 931
\bibitem{yngve1986} S. J. Yngve, \emph{Am. J. Phys.} \textbf{54} (1986) 1103
\bibitem{aquino2007} N. Aquino, G. Campoy and H. E. Montgomery Jr., \emph{Int. J. Quantum Chem.} \textbf{107} (2007) 1548.
\bibitem{pfaffenzeller1996} O. Pfaffanzeller, Report No. JUL-3281, Forschungszentrum, Germany (1996) 195.
\bibitem{dutt2001} R. Dutt, A. Mukherjee and Y. P. Varshni, \emph{Phys. Lett. A} \textbf{280} (2001) 318.
\bibitem{capitelli2009} M. Capitelli and D. Giordano, \emph{Phys. Rev. A} \textbf{80} (2009) 032113.
\bibitem{soullard2004} J. Soullard, R. Santamaria and S.A. Cruz, \emph{Chem. Phys. Lett.} \textbf{391} (2004) 187.
\bibitem{connerade1999} J.P. Connerade, V.K. Dolmatov, P.A. Lakshmi and S.T. Manson, \emph{J. Phys. B:
At. Mol. Opt. Phys.} \textbf{32} (1999) L239.
\bibitem{belosludov2003} V.R. Belosludov, T.M. Inerbaev, R.V. Belosludov and Y. Kawazoe, \emph{Phys. Rev.
B} \textbf{67} (2003) 155410.
\bibitem{decleva1999} P. Decleva, G. De Alti, G. Fronzoni and M.J. Stener, \emph{J. Phys. B: At. Mol. Opt.
Phys.} \textbf{325} (1999) 4523.  
\bibitem{hernandez1998} J. Hernandez-Rojas, J. Breton and J.M. Gomez Llorente, \emph{J. Chem. Phys.} \textbf{108}
(1998) 3498.
\bibitem{qiao2002} H.X. Qiao, T.Y. Shi and B.W. Li, \emph{Commun. Theor. Phys.} \textbf{37} (2002) 221.
\bibitem{connerade2001} J.P. Connerade, A.G. Lyalin, R. Semaoune, S.K. Semenov and A.V. Solovyov,
\emph{J. Phys. B: At. Mol. Opt. Phys.} \textbf{34} (2001) 2505.
\bibitem{kang2007} S. Kang, Q. Liu, H. Y. Meng and T. Y. Shi, \emph{Phys. Lett. A} \textbf{360} (2007) 608.
\bibitem{cremaschi1985} P. Cremaschi and J. L.Whitten, \emph{Surf. Sci.} \textbf{149(1)} (1985) 273-284
\bibitem{sakaguchi1995} H. Sakaguchi, T. Suenobu, K. Moriuchi, M. Yamagami, T. Yamaguchi and G. Anachi, \emph{J. Alloys Compd.} \textbf{221(1)} (1995) 212-217 
\bibitem{schlapbach2001} L. Schlapbach and A. Z\"{u}ttel, \emph{Nature} \textbf{414} (2001) 353-358.
\bibitem{ting2000} S. Ting-yun, Q. Hao-xue and L. Bai-wen, \emph{J. Phys. B: At. Mol. Opt. Phys.} \textbf{33} (2000) L349. 
\bibitem{micca2015} G. Micca Longo, S. Longo and D. Giordano, \emph{Phys. Scr.} \textbf{90} (2015) 025403.
\bibitem{foulkes2001} W. M. C. Foulkes, L. Mitas, R. J. Needs and G. Rajagopal, \emph{Rev. Mod. Phys.} \textbf{73} (2001) 33.
\bibitem{thijssen} J. M. Thijssen, \emph{Computational Physics}, Cambridge Univerity Press (1999).
\bibitem{reynolds1982} P. J. Reynolds, D. M. Ceperley, B. J. Alder and W. A. Lester Jr., \emph{J. Chem. Phys.} \textbf{77} (1982) 5593.
\bibitem{anderson1975} J. B. Anderson, \emph{J. Chem. Phys.} \textbf{63} (1975) 1499.
\bibitem{anderson1976} J. B. Anderson, \emph{J. Chem. Phys.} \textbf{65} (1976) 4121.
\bibitem{moskowitz1982} J. W. Moskowitz, K. E. Schmidt, M. A. Lee and M. H. Kalos, \emph{J. Chem. Phys.} \textbf{77} (1982) 349.
\bibitem{aquino2005} N. Aquino, G. Campoy and A. Flores-Riveros, \emph{Int. J. Quantum Chem.} \textbf{103} (2005) 267.
\bibitem{burns1993} R. G. Burns,  \emph{Mineralogical applications of crystal field theory (Vol. 5)} (1993)  Cambridge University Press.
\bibitem{cassidy2011} D. B. Cassidy, M. W. J. Bromley, L. C. Cota, T. H. Hisakado, H. W. K. Tom, and A. P. Mills, Jr., \emph{Phys. Rev. Lett.} \textbf{106} (2011) 023401.

\end{thebibliography}
\end{document}